\documentclass[journal=nalefd,manuscript=article]{achemso}

\usepackage{chemformula} 
\usepackage[T1]{fontenc} 
\usepackage{graphicx}
\usepackage{epstopdf}
\usepackage{dcolumn}
\usepackage{bm}
\usepackage{color}
 \usepackage{array,multirow,graphicx}

\author{Elena Voloshina}
\affiliation[SHU]
{Department of Physics, Shanghai University, 200444 Shanghai, China}
\email{voloshina@shu.edu.cn}
\alsoaffiliation[FUB]
{Institut f\"ur Chemie und Biochemie, Freie Universit\"at Berlin, Arnimallee 22, 14195 Berlin, Germany}
\author{Beate Paulus}
\affiliation[FUB]
{Institut f\"ur Chemie und Biochemie, Freie Universit\"at Berlin, Arnimallee 22, 14195 Berlin, Germany}
\author{Yuriy Dedkov}
\affiliation[SHU]
{Department of Physics, Shanghai University, 200444 Shanghai, China}
\email{dedkov@shu.edu.cn}

\title[]
  {Graphene Layer Morphology as an Indicator of the Metals Alloy Formation at the Interface}


\begin{document}




\begin{abstract}

The intercalation of different species in graphene-metal interfaces is widely used to stabilise the artificial phases of different materials. However, formation of the surface alloys upon the guest-metal intercalation is still an open question, which is very important for the fabrication of graphene-based interfaces with desired properties. Here, the widely studied interfaces of graphene with Ru(0001) and Ir(111) were modified using intercalation of a thin Mn layer and investigated by means of scanning tunnelling microscopy (STM) accompanied by density functional theory (DFT) calculations, which reproduce the observed experimental data. It is found that Mn forms a pseudomorphic layer on Ru(0001) under a strongly buckled graphene layer. In case of Mn intercalation in graphene/Ir(111), a buried thin layer of MnIr alloy is formed beneath the first Ir layer under a flat graphene layer. This unexpected observation is explained on the basis of phase diagram pictures for the Mn-Ru and Mn-Ir systems as well as via comparison of calculated total energies for the respective interfaces. Our results shed light on the understanding of mechanisms of the alloys formation at the graphene-metal interfaces and demonstrate their potential for the preparation of tailored interfaces for future graphene-based applications.

\end{abstract}

\section*{}

Graphene (gr) layers on metallic supports are in the focus of solid state chemistry and physics and materials science for many years~\cite{Tontegode:1991ts,Oshima:1997ek,Batzill:2012,Dedkov:2015kp,Yang:2020fda}. These works, initially motivated by studies of the catalytic activity of transition metal surfaces, received increased attention after discovery of the fascinating properties of graphene in 2004~\cite{Novoselov:2005es,Zhang:2005gp}. Following these discoveries of the transport properties of graphene, the renewed interest to graphene-metal systems led to several interesting findings. For example, it was shown that huge single- and bi-layer-thick graphene sheets can be synthesised on polycrystalline metal foils and then transferred on the desired substrate for further applications (with some drawbacks)~\cite{Bae:2010,Ryu:2014fo}; it was proposed that graphene moir\'e structures on close-packed surfaces of $4d$ and $5d$ metals can be used for fabrication of ordered arrays of clusters~\cite{NDiaye:2009a,MartinezGalera:2014hs,DiezAlbar:2019kq}, which are useful for the fundamental studies of catalytic properties of single clusters. Initially studied protective properties of graphene on metals, where graphene is considered as an inhibitor for surface reactions~\cite{Dedkov:2008d,Dedkov:2008e,Sutter:2010bx,Weatherup:2015cx}, were extended to the studies of effects of confined catalysis at the graphene-metal interface where, for example, the quantum effects due the space confinement at the graphene-Pt(111) interface promote the CO-oxidation~\cite{Yao:2014hy,Shifa:2019gb}. 

One of the interesting areas in the studies of graphene-metal interfaces is the possibility to modify the structural, electronic and magnetic properties of graphene via intercalation of different species. Here, the range of materials is spread from metallic elements, like Cu, Ag, Au or Fe~\cite{Dedkov:2001,Dedkov:2008e,Varykhalov:2010a} to gaseous or molecular species~\cite{Shikin:2000a,Granas:2013tl,Dedkov:2017jn}. Although, the mechanism of intercalation in graphene-based interfaces is not fully clear and different pathways for the species penetration were discussed in the literature~\cite{Emtsev:2011fo,Petrovic:2013vz,Vlaic:2014gx} without any theoretical support, this allows to create artificial graphene-metal interfaces with interesting structural, electronic and magnetic properties. However, for example, in the case of metals intercalation in graphene-metal interface, the formation of sharp metallic interfaces vs. surface (interfaces) alloying is always an open question during such studies, because in most cases the used experimental methods do not give a clear answer.

Here, we present a combined scanning tunnelling microscopy (STM) and density functional theory (DFT) study of Mn intercalation in the gr/Ru(0001) and gr/Ir(111) systems. Both parent systems present representative examples of strongly- and weakly buckled graphene-metal interfaces, respectively, which demonstrate strongly different behaviour for the interface formation after Mn intercalation. Despite the expected pseudomorphic growth of Mn on Ru(0001) and Ir(111) under graphene with the formation of strongly buckled graphene above, we found that the formation of the metallic interface is different for both cases. Whereas for gr-Mn-Ru(0001) the expected behaviour, with the formation of a strongly buckled graphene layer, is confirmed by STM and DFT, the gr-Mn-Ir(111) system surprisingly demonstrates the formation of flat graphene on top. Such behaviour in the later case is explained by the formation of a buried Mn-Ir alloy underneath the gr/Ir interface bi-layer. These findings are confirmed by DFT calculations and very good agreement is found between experimental and theoretical data.

\begin{figure}
\center
\includegraphics[width=0.8\columnwidth]{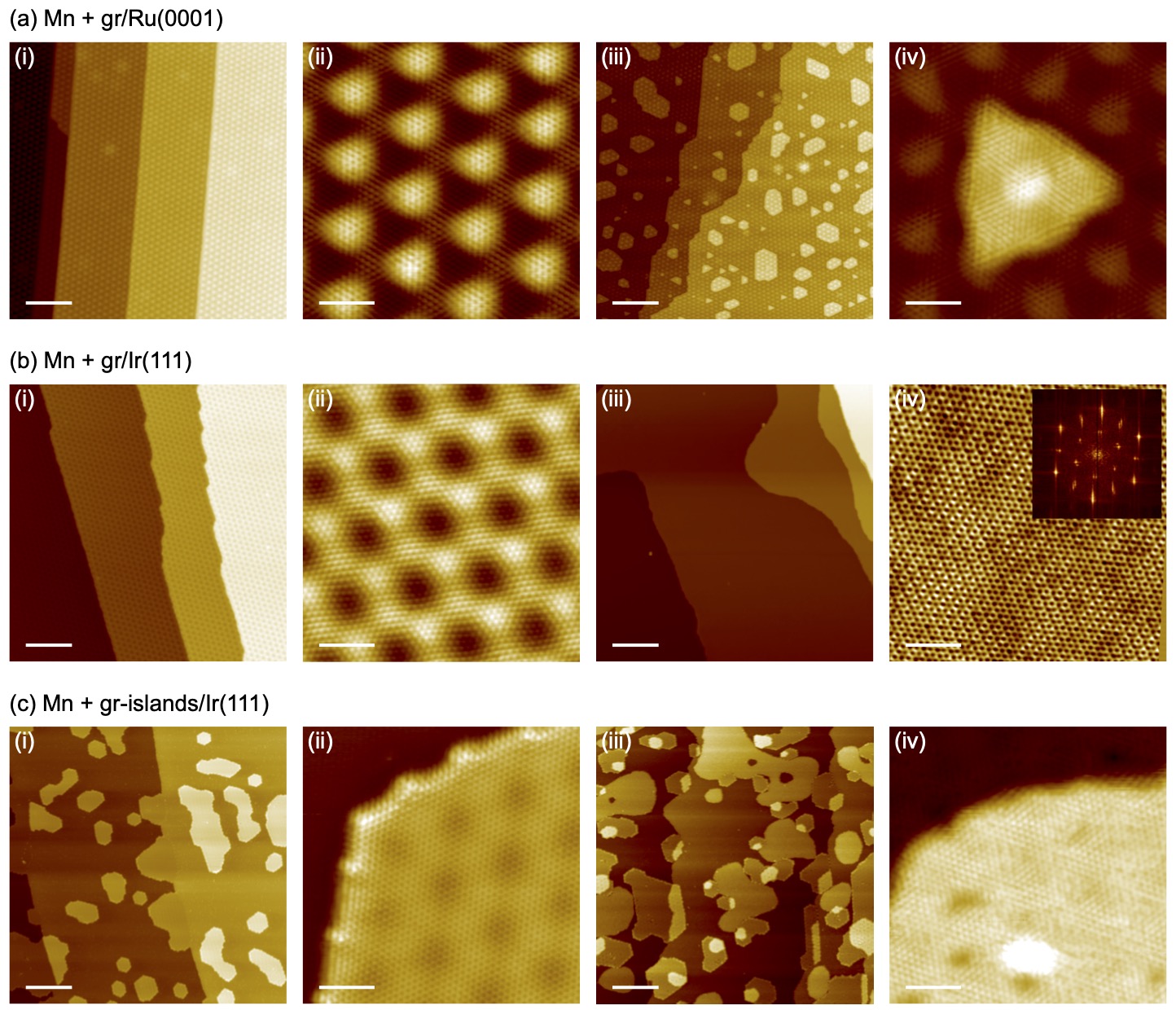}\\
\vspace{1cm}
\caption{STM images of Mn intercalation in gr/Ru(0001) and gr/Ir(111). (i) and (iii) are large scale STM images of parent systems and systems after Mn intercalation, respectively; (ii) and (iv) are zoomed atomically-resolved STM images of the same systems, respectively. (a) Mn+gr/Ru(0001): (i) $120\times120\,\mathrm{nm}^2$, $U_T=+300\,\mathrm{mV}$, $I_T=1\,\mathrm{nA}$, (ii) $10\times10\,\mathrm{nm}^2$, $U_T=+50\,\mathrm{mV}$, $I_T=3\,\mathrm{nA}$, (iii) $120\times120\,\mathrm{nm}^2$, $U_T=+300\,\mathrm{mV}$, $I_T=1\,\mathrm{nA}$, (iv) $10\times10\,\mathrm{nm}^2$, $U_T=+100\,\mathrm{mV}$, $I_T=1\,\mathrm{nA}$; (b) Mn+gr/Ir(111): (i) $120\times120\,\mathrm{nm}^2$, $U_T=+300\,\mathrm{mV}$, $I_T=1\,\mathrm{nA}$, (ii) $10\times10\,\mathrm{nm}^2$, $U_T=-50\,\mathrm{mV}$, $I_T=2\,\mathrm{nA}$, (iii) $120\times120\,\mathrm{nm}^2$, $U_T=+500\,\mathrm{mV}$, $I_T=1\,\mathrm{nA}$, (iv) $10\times10\,\mathrm{nm}^2$, $U_T=-50\,\mathrm{mV}$, $I_T=10\,\mathrm{nA}$ (inset shows the respective FFT image); (c) Mn+gr-islands/Ir(111): (i) $120\times120\,\mathrm{nm}^2$, $U_T=+300\,\mathrm{mV}$, $I_T=1.6\,\mathrm{nA}$, (ii) $10\times10\,\mathrm{nm}^2$, $U_T=+100\,\mathrm{mV}$, $I_T=1.6\,\mathrm{nA}$, (iii) $120\times120\,\mathrm{nm}^2$, $U_T=+300\,\mathrm{mV}$, $I_T=1.2\,\mathrm{nA}$, (iv) $10\times10\,\mathrm{nm}^2$, $U_T=-100\,\mathrm{mV}$, $I_T=15\,\mathrm{nA}$. White scale bars in (i,iii) and (ii,iv) correspond to $20$\,nm and $2$\,nm, respectively.}
\label{fig:Mn_and_grIrRu_LargeScaleZoom}
\end{figure}

Figure~\ref{fig:Mn_and_grIrRu_LargeScaleZoom} compiles the results on the Mn intercalation in gr/Ru(0001) and gr/Ir(111) (full graphene layer and graphene islands). Images in columns (i) and (ii) are large scale and atomically resolved STM images, respectively, of the parent systems. Both graphene-metal interfaces, gr/Ru(0001) and gr/Ir(111), are characterised by the relatively large mismatch between graphene and the metal lattices, that leads to the formation of so-called periodic moir\'e structures, clearly resolved in STM images as long wave modulations of the atomically-resolved imaging contrast~\cite{Dedkov:2015iza} (Fig.~\ref{fig:Mn_and_grIrRu_LargeScaleZoom}, column ii). Here, several high symmetry positions for adsorption of carbon atoms in a graphene layer on close-packed metallic surfaces can be identified. They are correspondingly called with respect to the adsorption position on metallic surface surrounded by a carbon ring -- ATOP, HCP, FCC~\cite{Voloshina:2012c,Dedkov:2015kp}, with the ATOP position having the largest distance between graphene and metallic surface. In STM images collected at small bias voltages ($U_T<|\pm1|$\,V), graphene on Ru(0001) is imaged in the \textit{direct} contrast with ATOP place as a bright spot~\cite{Stradi:2011be,Voloshina:2016jd}, whereas graphene on Ir(111) is imaged in the \textit{inverted} contrast with ATOP place as a dark spot~\cite{Voloshina:2013dq}, that is explained by the formation of the respective interface electronic states in the vicinity of the Fermi level.

In our experiments Mn was evaporated on gr/Ru(0001) and gr/Ir(111) from an e-beam source and in both cases this leads to the formation of the ordered arrays of Mn clusters on top of a graphene layer (see Fig.\,S1 in Supplementary Material for summary). Presented in Fig.\,S1 STM images for the systems with Mn clusters formed on a graphene moir\'e structure were collected at small bias voltages, which are usually used to obtain atomic resolution of a graphene lattice. The absence of the clear atomic resolution in the presented small-scale STM data (Fig.\,S1 in Supplementary Material) confirms the formation of ordered arrays of Mn clusters in the considered systems. The careful and systematic analysis of the available experimental data allows to conclude that Mn clusters are adsorbed at the HCP (C$^{top}$-C$^{fcc}$, carbon atoms are located above interfacial Ru atoms and above \textit{fcc} hollow sites) high-symmetry positions of gr/Ru(0001) and gr/Ir(111). In case of the adsorption of Mn clusters on graphene-islands on Ir(111), one can find that the Mn-coverage strongly depends on the islands' alignment on the Ir(111) surface and correspondingly on the periodicity of the graphene moir\'e lattice. As it was previously shown, this effect leads to different adsorption energies for metallic clusters on the graphene-metal system~\cite{Sutter:2012kb,Zhang:2020ba} and for some angles (moir\'e periodicities) graphene remains uncovered with metal atoms.

Annealing of the Mn/gr/Ru(0001) and Mn/gr/Ir(111) systems at $T_a=500^\circ$\,C for $15$\,min leads to the penetration of Mn atoms underneath the graphene layer (Fig.~\ref{fig:Mn_and_grIrRu_LargeScaleZoom}(iii,iv) and Fig.\,S2 in Supplementary Material). However, the systems obtained after Mn intercalation demonstrate strong difference between them in the resulting morphology, although in both cases the formation of a strongly corrugated graphene layer is expected after intercalation. This expectation is based on the previous experimental results on the intercalation of the open $3d$-shell metals, like Fe, Co, Ni in gr/Ru and gr/Ir interfaces~\cite{Liao:2012jw,Pacile:2013jc,Decker:2013ch,Bazarnik:2013gl,Vlaic:2018fg,Zhao:2018gh}. In all referenced cases a sharp interface between graphene and intercalated material is formed, where atoms of the intercalant are pseudomorphically arranged on close-packed surfaces of Ru(0001) or Ir(111). The resulting corrugation of a graphene layer in such systems is more than $1$\,\AA\ with spatially modulated interaction strength between graphene and the underlying $3d$ metal~\cite{Voloshina:2014jl,Dedkov:2015kp}.  

After Mn intercalation in gr/Ru(0001) a strongly corrugated graphene layer is formed above Mn-Ru(0001), whereas a relatively flat graphene is formed on top of Mn-Ir(111) (Fig.~\ref{fig:Mn_and_grIrRu_LargeScaleZoom}(iii,iv) and Fig.\,S2 in Supplementary Material). The corrugation of the graphene layers in these systems as extracted from STM images is $1$\,\AA\ and $0.15$\,\AA\ for gr/Mn-Ru(0001) and gr/Mn-Ir(111), respectively. Moreover, a \mbox{$(2\times2)$} superstructure for the gr/Mn-Ir(111) system is clearly resolved in the STM images and confirmed by the corresponding Fast Fourier Transformation (FFT) analysis (see respective inset in Fig.~\ref{fig:Mn_and_grIrRu_LargeScaleZoom}). Taking into account all observations and facts for both considered graphene-based interfaces, we can conclude that sharp interfaces between graphene, Mn layer and Ru(0001) support are formed for the gr/Mn-Ru(0001) system. In case of gr/Mn-Ir(111) the situation is not so simple, requiring additional analysis.

\begin{figure}
\center
\includegraphics[width=0.8\columnwidth]{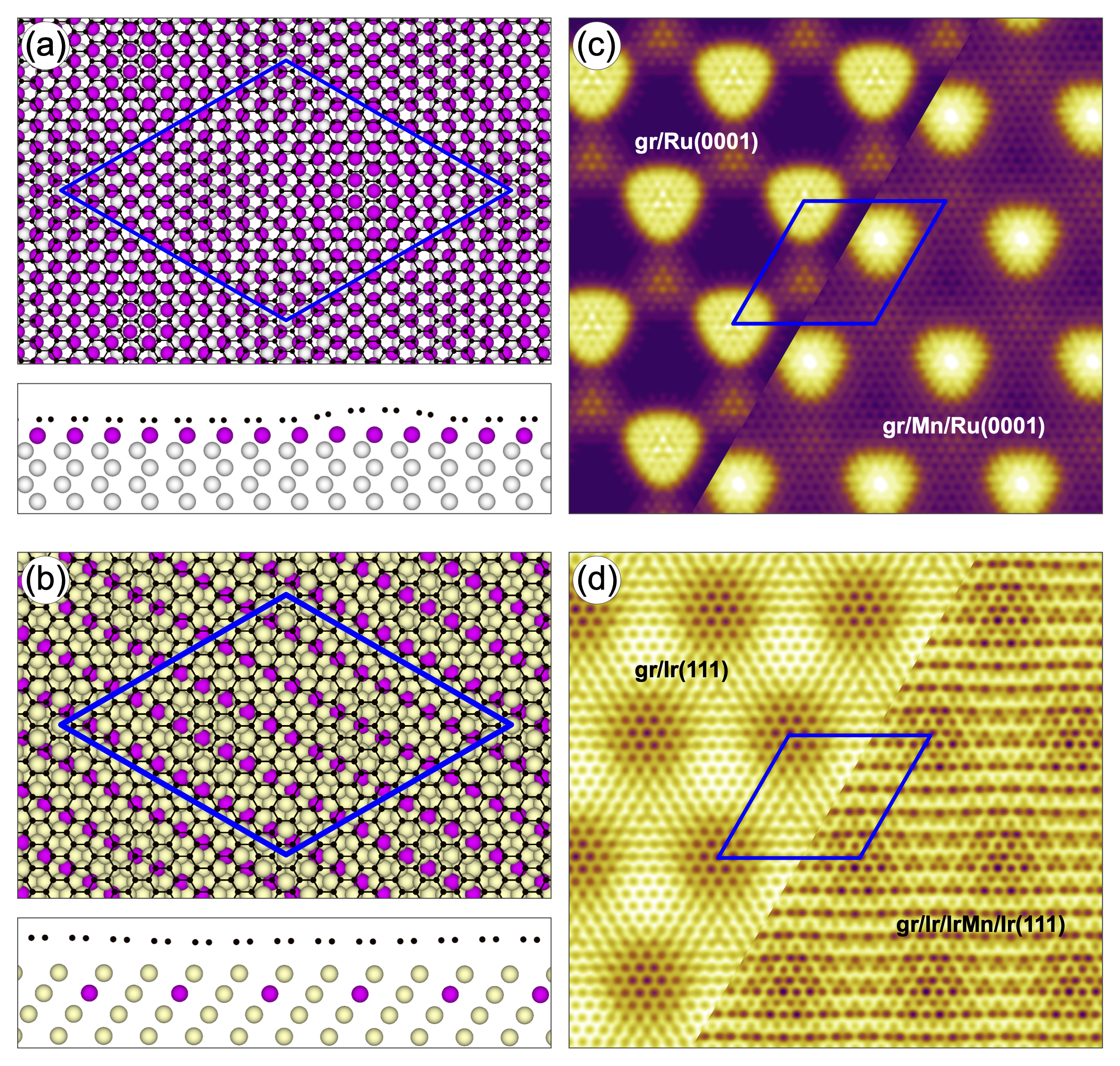}\\
\vspace{1cm}
\caption{Theoretical results obtained for Mn-intercalated graphene-metal interfaces, gr/Mn-Ru(0001) and gr/Mn-Ir(111). (a) Side view of the DFT optimised most energetically favourable gr/Mn-Ru(0001) interface, where the Mn layer is sandwiched between the top graphene layer and the Ru(0001) substrate. (b) Side view of the DFT optimised of one of the most energetically favourable gr/Mn-Ir(111) interfaces, where 1\,ML of the IrMn ordered alloy is buried beneath the top Ir layer under a graphene layer. (c) Theoretically computed STM images of gr/Ru(0001) and gr/Mn/Ru(0001). (d) Theoretically computed STM images of gr/Ir(111) and gr/Ir/IrMn/Ir(111). Blue rhombuses mark the respective unit cells for considered structures. In (c) and (d) the integration of the electronic states around $E_F$ was performed for the energy range corresponding to the experimental bias voltage of $U_T=-300$\,mV and the obtained images correspond to the experimental tunnelling current of $I_T\approx1.2$\,nA.}
\label{fig:DFT_results}
\end{figure}

In order to understand the observed effects, we performed large-scale DFT calculations for different Mn-intercalation systems. In the first case, $1$\,ML-Mn is placed between graphene and the close-packed metallic support, Ru(0001) or Ir(111) (Fig.~\ref{fig:DFT_results}(a)), while in the second case $1$\,ML-Mn or $1$\,ML of the Mn-Ir(Ru) alloy is buried below the interface Ir(Ru) layer under graphene (Fig.~\ref{fig:DFT_results}(b)). In the later case, ordered and disordered alloys were considered. As a criteria for the successful theoretical modelling of the experimental data, the total energy for two concurrent systems as well as agreement between experimental and theoretical STM images (particularly, the extracted graphene corrugation) are taken into account. Although, the alloying of the metallic intercalant with a graphene support is always discussed in graphene-metal related studies, this problem was not studied in details. Particularly, rare experimental and theoretical works always consider the formation of a surface alloy between atoms of intercalant and graphene support~\cite{Drnec:2015kn,Brede:2016fq}, which explains the obtained experimental data in these works, especially STM images. As it will be discussed further, although sharp interfaces are formed in the case of gr/Mn-Ru(0001), the earlier discussed considerations do not support the available experimental data for the Mn intercalation in gr/Ir(111). Table\,S3 in the Supplementary Material summarises all information obtained from the DFT calculations (size of the unit cell, number of atoms in the system, total energies, corrugation of graphene, etc.) and used below for analysis of the observed effects.

\begin{figure}
\center
\includegraphics[width=0.75\columnwidth]{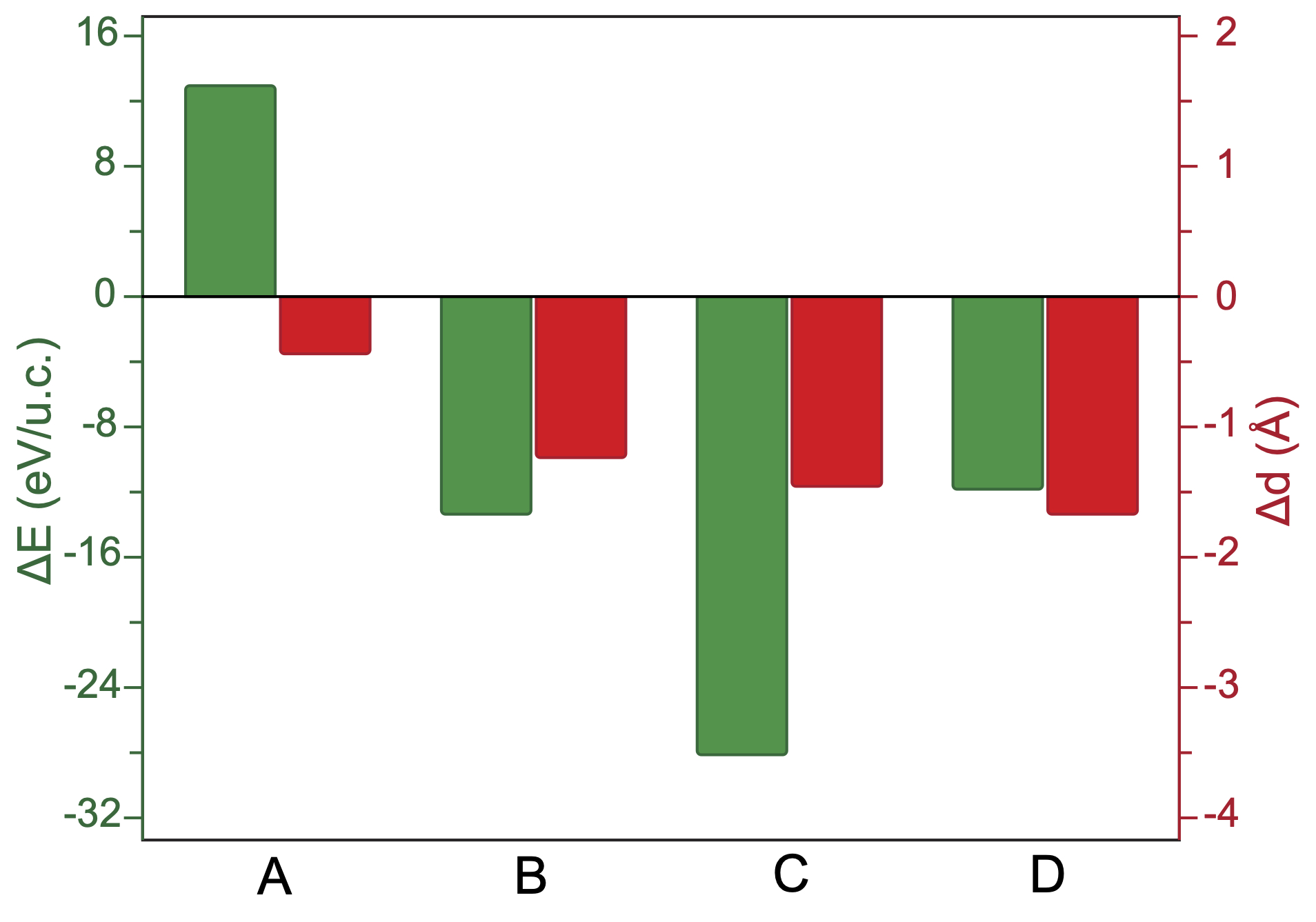}\\
\vspace{1cm}
\caption{Comparison of the total energies and graphene corrugations for the considered Mn-intercalated interfaces. Green and red bars show the change of the total energy of the system and the graphene corrugation, respectively, upon placing the respective Mn-containing layer from the gr/Mn-M/M state to the buried state - gr/M/Mn-M/M (M = Ru, Ir). Zero values for the change of the total energy and graphene corrugation correspond to the respective absolute values from Table\,S3 for the gr/Mn-M/M state. Considered systems: A - gr/Mn/Ru(0001), B - gr/Mn/Ir(111), C - gr/IrMn-ordered/Ir(111), D - gr/IrMn-disordered/Ir(111).}
\label{fig:DFT_results_energies}
\end{figure}

For the gr/Mn-Ru(0001) system, the DFT calculated total energy difference between gr/Ru/Mn/Ru(0001) and gr/Mn/Ru(0001) and is $+13.022$\,eV (Fig.~\ref{fig:DFT_results_energies}), which corresponds to $+12$\,meV per atom in the considered system (assuming the atoms alignment adopted from the experimental STM data: gr$^{ATOP}_{gr/Ru}$$\rightarrow$gr$^{HCP}_{gr/Mn/Ru}$ and gr$^{HCP}_{gr/Ru}$$\rightarrow$gr$^{ATOP}_{gr/Mn/Ru}$; C-atoms positions are taken with respect to the underlying metal slab). This result immediately shows that the gr/Mn/Ru(0001) system is formed during Mn intercalation in gr/Ru(0001). In this system graphene has a corrugation of $1.519$\,\AA\ with a minimal distance between graphene and Mn layer of $1.861$\,\AA, i.\,e. the morphology of the gr/Mn/Ru(0001) system obtained in our DFT calculations is similar to the one of the parent gr/Ru(0001) system (cf. $1.302$\,\AA\ and $2.123$\,\AA\ for corrugation and graphene-Ru distance, respectively). This result clearly supports the experimental observation, where the morphology of graphene is very similar for gr/Ru(0001) and gr/Mn/Ru(0001) areas in the STM images (Fig.~\ref{fig:Mn_and_grIrRu_LargeScaleZoom}). The respective calculated STM images for gr/Ru(0001) and gr/Mn/Ru(0001) are shown in Fig.~\ref{fig:DFT_results}(c).  

In case of the gr/Mn-Ir(111) intercalation system the situation is opposite. The first assumption to describe this system by the formation of gr/Mn/Ir(111) with the Mn layer pseudomorphically arranged on Ir(111) and sharp interfaces between layers is not supported by the calculated equilibrium crystallographic structure. In such a structure graphene is strongly buckled with its corrugation of $1.525$\,\AA\ and minimal distance between graphene and Mn of $1.798$\,\AA. Thus such a structure is very similar to previously studied gr/Ni/Ir(111)~\cite{Pacile:2013jc}, gr/Co/Ir(111)~\cite{Decker:2013ch}, and gr/Mn/Ru(0001) (see above). This result is in contradiction to the observation from the STM experiments, where a relatively flat graphene layer was observed for the gr/Mn-Ir(111) system with a corrugation of only $0.15$\,\AA. The attempt to improve this situation via insertion of a Mn$_x$Ir$_y$ alloy (ordered and disordered) between graphene and Ir(111) does not lead to an acceptable result -- graphene remains strongly buckled with its corrugation of more than $1.8$\,\AA\ (see Table\,S3 in Supplementary Material), which is in strong contradiction with experimental results.

The significant improvement is only achieved when buried Mn or MnIr monolayers are considered for the modelling of the gr/Mn-Ir(111) intercalation systems. In this case the calculated total energies for these systems are significantly lower compared to those for the systems, where Mn or MnIr are placed directly underneath a graphene layer (see Table\,S3 in the Supplementary Material). The lowering of the total energy is ranged between $19.6$\,meV and $38.1$\,meV per atom in the considered systems (corresponds to $11.888$\,eV and $28.272$\,eV (Fig.~\ref{fig:DFT_results_energies}) for the systems consisting of more than 600 atoms; see Table\,S3 in the Supplementary Material for detailed structures). Also the graphene layer becomes relatively flat with its corrugation varied between $0.277$\,\AA\ and $0.429$\,\AA, which is comparable with the one of $0.358$\,\AA\ for the parent gr/Ir(111) interface. Taking into account the fact that during STM measurement the local electronic structure is probed and graphene corrugation is measured indirectly, one can conclude that a rather good agreement between experiment and theory is achieved in this case. Also comparing the experimental and calculated STM images, the formation of an ordered MnIr alloy buried below interface Ir layer under graphene is concluded. The respective calculated STM images for gr/Ir(111) and gr/Ir/IrMn/Ir(111) are shown in Fig.~\ref{fig:DFT_results}(d) and the FFT analysis of the calculated STM image of the later system also clearly indicates the existence of a $(2\times2)$ periodicity in these data (see Fig.~S4 in the Supplementary Material).

The above presented picture for gr/Mn-Ru(0001) and gr/Mn-Ir(111) is supported by the general consideration on the formation of surface and sub-surface (buried) alloys. The difference in the formation of the (A+B)/A and A/(A+B)/A systems is connected to the so-called segregation energy, $E_{seg}$, which can be calculated as the difference between the total energies of the system with the impurity in a surface layer and in the bulk~\cite{Ruban:1999kq,Ruban:1999kqa} (A is a close-packed host (Ru or Ir) and B is a solute (Mn)). If $E_{seg}<0$ then the surface alloy can be formed [i.\,e. (A+B)/A] and if $E_{seg}>0$ then the formation of the sub-surface alloy prevails [i.\,e. A/(A+B)/A]. The theoretical values of $E_{seg}$ for Mn-Ru and Mn-Ir are $-0.40$\,eV and $+0.09$\,eV, respectively, indicating the formation of sub-surface MnIr alloy for the gr/Mn-Ir(111) system~\cite{Okamoto:1996ku}. In case of the gr/Mn-Ru(0001) system the alloying of two metallic components is unfavourable according to their phase diagram~\cite{Hellawell:1959gf}, thus leading to the formation of the well-ordered gr/Mn/Ru(0001) intercalation system with sharp interfaces between layers (see Fig.\,S5 in Supplementary Material for Mn-Ru and Mn-Ir phase diagrams~\cite{Raub:1955aa}). Similar observations were also made for the gr/Mn-Rh(111) and hBN/BN-Rh(111) systems ($E_{seg}=-0.08$\,eV for Mn-Rh~\cite{Ruban:1999kq,Ruban:1999kqa}), where Mn was found in the surface region directly underneath a graphene layer~\cite{Zhang:2013bw}. However, a clear discrimination between formation of a single Mn layer or a MnRh surface alloy was not made, although phase diagram analysis favours the later case~\cite{Raub:1955aa}.

In summary, we studied the intercalation of thin Mn layers in the gr/Ru(0001) and gr/Ir(111) interfaces using systematic STM and DFT approaches. Our results unequivocally demonstrate different final results for both systems. While for the gr/Mn/Ru(0001) system the formation of sharp interfaces between all components is found, the intercalation of Mn in gr/Ir(111) can lead to the formation of a sub-surface (buried) alloy below and Ir interface layer underneath graphene. These findings are understood on the basis of large-scale DFT calculations giving a significant lowering of the total energy for the system with the buried MnIr layer compared to the one with a surface MnIr alloy under a graphene layer. These results are also supported by the general thermodynamical considerations of phase diagrams for the respective binary systems as well as general theoretical calculations on the impurities segregations in different close-packed metallic hosts. With these new results for the graphene-intercalation systems, we also suggest that additional structural spectroscopical studies of these or similar systems are performed for further revision of the previous experimental data. Our findings shed a light on one of the main questions rising in the studies of metal intercalation in graphene-based systems and are of paramount importance for the understanding of the structure and electronic properties of graphene-support interfaces. This knowledge will help in the synthesis of the desired interfaces for future graphene-based applications.


\section*{Experimental}
\paragraph{STM measurements.}
The STM measurements were performed in constant current modes at room temperature with an SPM Aarhus 150 equipped with a KolibriSensor from SPECS with a Nanonis Control system. In these measurements a sharp W-tip was used which was cleaned \textit{in situ} via Ar$^+$-sputtering. In the presented STM images the tunnelling bias voltage, $U_T$, is applied to the sample and the tunnelling current, $I_T$, is collected through the tip, which is virtually grounded. Tunnelling current and voltage values are given in the figure captions. The base pressure in the experimental station is below $8\times10^{-11}$\,mbar. Graphene layers were prepared on Ru(0001) and Ir(111) using C$_2$H$_4$ as a carbon precursor according to the recipes given in Refs.~\citenum{Voloshina:2016jd,Voloshina:2013dq}, respectively. Intercalation of Mn was performed via e-beam deposition and subsequent annealing of thin manganese layers (with thickness around $1$\,ML) on top of graphene layers. The respective annealing temperatures are noted in the text.

\paragraph{DFT calculations.}
DFT calculations based on plane-wave basis sets of $400$\,eV cut-off energy were performed with the Vienna \textit{ab initio} simulation package (VASP)~\cite{Kresse:1994cp,Kresse:1996kg}. The Perdew-Burke-Ernzerhof (PBE) exchange-correlation functional~\cite{Perdew:1996abc} was employed. The electron-ion interaction was described within the projector augmented wave (PAW) method~\cite{Blochl:1994fq} with C ($2s$, $2p$), Ir ($5d$, $6s$), Ru ($4d$, $5s$), Mn ($3d$, $4s$) states treated as valence states. The Brillouin-zone integration was performed on $\Gamma$-centred symmetry reduced Monkhorst-Pack meshes using a Methfessel-Paxton smearing method of first order with $\sigma = 0.2$\,eV. The $k$ mesh for sampling the supercell Brillouin zone are chosen to be at least as dense as $30 \times 30$, when folded up to the simple graphene unit cell. Dispersion interactions were considered by adding a $1/r^6$ atom-atom term as parameterised by Grimme (``D2'' parameterisation)~\cite{Grimme:2006fc}. The systems studied in the present work were considered in the supercell geometry due to the relatively lattice sizes mismatch between graphene and underlying metal. Each of such supercell is constructed from a slab of five layers of metal, a graphene layer adsorbed on one (top) side of a metal slab and a vacuum region of approximately $20$\,\AA. The lattice constant in the lateral plane was set according to the optimised value of bulk metal ($a_\mathrm{Ir(111)} = 2.723$\,\AA\ and $a_\mathrm{Ru(0001)}=2.703$\,\AA). The positions ($x$, $y$, $z$-coordinates) of C atoms and intercalant as well as $z$-coordinates of the two topmost layers of the substrate were fully relaxed until forces became smaller than $0.02$/eV\,\AA$^{-1}$. The STM images are calculated using the Tersoff-Hamann formalism~\cite{Tersoff:1985}.

\begin{acknowledgement}
The North-German Supercomputing Alliance (HLRN) is acknowledged for providing computer time.
\end{acknowledgement}

\begin{suppinfo}
The following files are available free of charge
\begin{itemize}
  \item Additional experimental and theoretical data (PDF) can be downloaded via link: https://pubs.acs.org/doi/10.1021/acs.jpclett.0c03271
\end{itemize}

\end{suppinfo}

\bibliography{references.bib}


\clearpage

\clearpage

\clearpage



\end{document}